\documentclass
[reprint,aps,prl,twocolumn,nobibnotes,superscriptaddress]{revtex4-1}%
\usepackage[utf8]{inputenc}
\usepackage{amsmath}
\usepackage{physics}
\usepackage{braket}
\usepackage{amssymb}
\usepackage{graphicx}
\usepackage{xcolor}
\usepackage[most]{tcolorbox}
\usepackage[unicode=true, breaklinks=false, pdfborder={0 0 1}, backref=false,
colorlinks=true, linkcolor=blue, urlcolor=blue, citecolor=blue]{hyperref}
\usepackage{amsfonts}%
\providecommand{\U}[1]{\protect\rule{.1in}{.1in}}
\setcitestyle{numbers,square}
\begin{document}

\title{Spin-Wave Doppler Shift by Magnon Drag in Magnetic Insulators}
\author{Tao Yu}
\affiliation{Max Planck Institute for the Structure and Dynamics of Matter, Luruper Chaussee 149, 22761 Hamburg, Germany}
\author{Chen Wang}
\affiliation{Center for Joint Quantum Studies and Department of Physics, School of Science, Tianjin University, Tianjin 300350, China}
\author{Michael A. Sentef}
\affiliation{Max Planck Institute for the Structure and Dynamics of Matter, Luruper Chaussee 149, 22761 Hamburg, Germany}
\author{Gerrit E. W. Bauer}
\affiliation{Institute for Materials Research $\&$ WPI-AIMR $\&$ CSRN, Tohoku University, Sendai 980-8577, Japan}
\date{\today}

\begin{abstract}
The Doppler shift of the quasiparticle dispersion by charge currents is
responsible for the critical supercurrents in superconductors and
instabilities of the magnetic ground state of metallic ferromagnets. Here we
predict an analogous effect in thin films of magnetic insulators in which
microwaves emitted by a proximity stripline generate coherent chiral spin
currents that cause a Doppler shift in the magnon dispersion. The spin-wave
instability is suppressed by magnon-magnon interactions that limit spin
currents to values close to but below the threshold for the instability. The
spin current limitations by the backaction of magnon currents on the magnetic
order should be considered as design parameters in magnonic devices.

\end{abstract}
\maketitle

\textit{Introduction.}---Realization of a large spin current is an
important pursuit in spintronics. Electrically insulating magnetic films are
promising candidate to achieve this goal, allowing low-dissipation information
processing by magnons
\cite{magnonics1,magnonics2,magnonics3,magnonics4,spin_insulatronics}. The
presently most suitable material to study magnon dynamics is yttrium iron
garnet (YIG), a ferrimagnet with high Curie temperature and arguably the
lowest damping \cite{low_damping_nanometer,saga_of_YIG}. Ultrathin YIG films
with thicknesses below 10~nm maintain very high magnetic quality
\cite{ultrathin_YIG,7nm} and a strongly enhanced Drude-type magnon
conductivity \cite{transistor,nonlinear_electric1,Duine} that should be
suitable to carry large spin currents. Recently, large spin currents were
observed in ultrathin YIG transistors with DC-current biased Pt gates that
inject a large number of nonequilibrium magnons
\cite{Suhl,turbulence,magnon_BEC,nonlinear_microwave} into the conducting
channel \cite{nonlinear_electric1,nonlinear_electric2}.

A Doppler shift of Bogoliubov quasiparticles under an electric current bias is
responsible for critical supercurrents in superconductors
\cite{SC_shift1,SC_shift2,SC_shift3}. Similar effect can happen in metallic
ferromagnets when using electric currents to excite magnetization dynamics by
the spin-transfer torque \cite{Shufeng_current,Tserkovnyak_current}. The
charge current induces a Doppler shift, i.e., a tilt of the spin-wave
dispersion of a homogeneous magnetization in momentum space, which could
trigger a spin-wave instability
\cite{instability_Doppler1,instability_Doppler2,Duine_Doppler} and modulate
the magnetic ground state \cite{Tatara}. These obviously do not apply to
magnetic insulators that cannot carry an electric charge current. However,
magnetic insulators are also conduits for (magnonic) spin currents that as
reported here cause a non-linear Doppler effect by magnon-magnon drag, which
also limit the spin current to a material dependent maximum.

\begin{figure}[t]
\begin{center}
{\includegraphics[width=8.2cm]{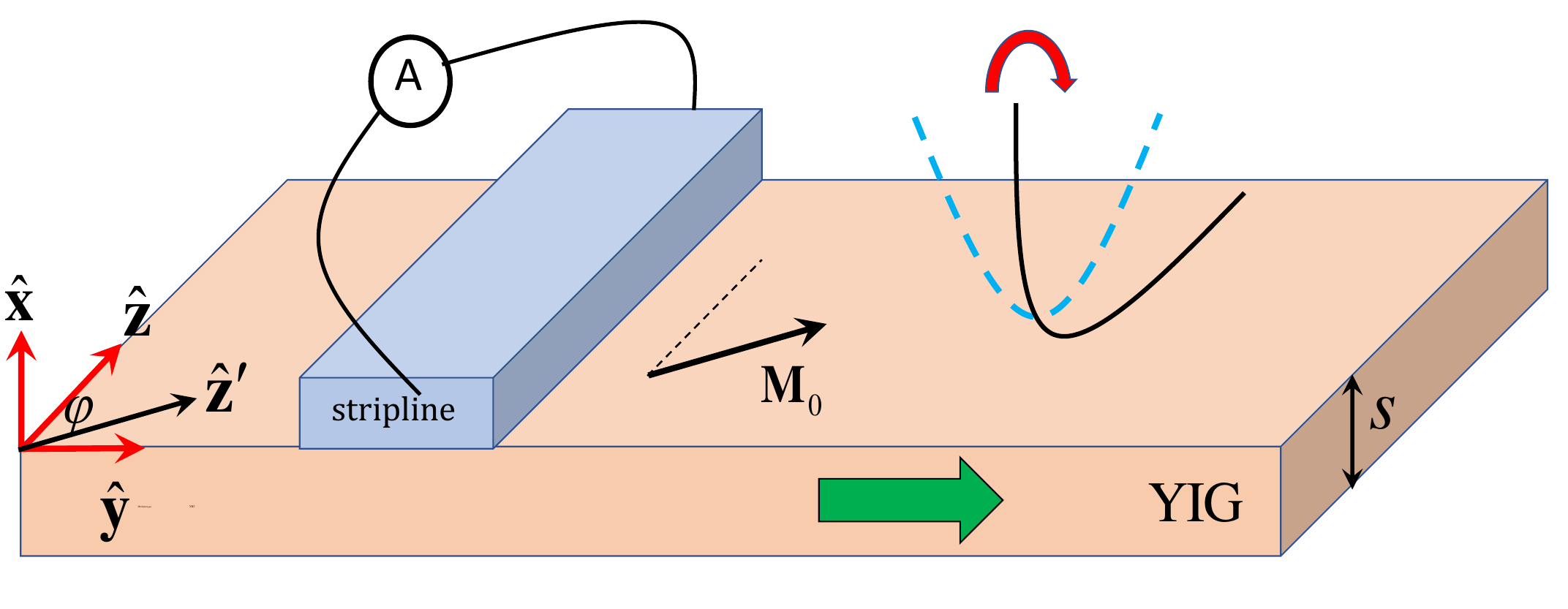}}
\end{center}
\caption{(Color online) Doppler effect of thin magnetic films driven
by pure magnon current. A long stripline along the $\hat{\mathbf{z}}
$-direction is illustrated to pump the magnon current (the green thick arrow)
in YIG films of thickness $s$ that causes the tilt of magnon dispersion, as
shown by the red thick arrow and parabolic bands. The in-plane magnetization
is saturated with a relative angle $\varphi$ to the stripline direction.}%
\label{nonlinear_YIG}%
\end{figure}

In this Letter, we formulate the dynamics of long-wavelength coherent magnons
of thin YIG films in the presence of large magnon currents that are pumped by
stripline microwaves as depicted in Fig.~\ref{nonlinear_YIG}. The
polarization-momentum locked AC magnetic field emitted by a microwave
stripline \cite{poineering_1,poineering_2,stripline_chirality} coherently
populates magnon states at one side of the stripline with a unidirectional
magnon current. We report here that (i) magnon interactions limit the
magnitude of this magnon current and the chirality of the pumping, and (ii) an
interaction-induced drag effect by the spin current on the magnon dynamics in
the form of a magnonic Doppler shift that tilts the spin-wave dispersion into
the current direction. The physics of the reported Doppler effect differs
strongly from the magnon-drag by phonon \cite{phonon} or electron
\cite{electron,instability_Doppler1,instability_Doppler2,Duine_Doppler}
currents. Its phenomenology is intriguingly similar to an interfacial
Dzyaloshinskii-Moriya interaction (DMI) \cite{DMI1,DMI2,DMI_Moon}, but can be
tuned by the excitation power. Interaction renders a linear (rather than
quadratic) dependence of the excited spin current amplitude at small driving
currents. When
the drive currents reach a critical value the Doppler shift leads to a
dispersion in which the magnon energy vanishes for a finite momentum state,
which corresponds to an instability of the ferromagnetic order. However, for
stronger drives, higher-order magnon interactions stabilize the magnetization
ground state and suppress the spin-wave instability by breaking the chirality
of chiral pumping. We thereby predict a maximum spin current that is close to but
(in the absence of a DMI assist) not large enough to cause a spin-wave
instability. 

\textit{Maximal spin current.}---We consider an in-plane magnetized YIG film
with thickness $s=\mathcal{O}(10)$~nm and saturated magnetization $M_{s}$,
with surface normal oriented along the $\hat{\mathbf{x}}$-direction. An
in-plane static magnetic field $\mathbf{H}_{\mathrm{app}}$ is applied at an
angle $\varphi$ to the stripline $\hat{\mathbf{z}}$-direction
(Fig.~\ref{nonlinear_YIG}). The Hamiltonian of the magnetic order reads
\begin{equation}
\hat{H}=\mu_{0}\int\left(  \frac{\alpha_{\mathrm{ex}}}{2}(\nabla
\hat{\mathbf{M}})^{2}-\hat{\mathbf{M}}\cdot\mathbf{H}_{\mathrm{app}}\right)
d\mathbf{r},
\end{equation}
where $\mu_{0}$ is the vacuum permeability, $\alpha_{\mathrm{ex}}$ is the
exchange stiffness, and $\mathbf{M}$ is the magnetization. We disregard
anisotropies \cite{dipolar_exchange_1,dipolar_exchange_2} because the crystal
ones are small in YIG, while the dipolar ones are strongly suppressed in the
thin film limit \cite{Bayer,chiral_excitation}. The exchange length in YIG is
$\lambda_{\mathrm{ex}}=2\pi\sqrt{\alpha_{\mathrm{ex}}}=109$~nm since
$\alpha_{\mathrm{ex}}=3\times10^{-16}$~m$^{2}$ \cite{Klingler,one_wire}. The
magnetization dynamics then obeys a Landau-Lifshitz-Gilbert (LLG) equation
\begin{equation}
\frac{d\mathbf{M}}{dt}=-\mu_{0}\gamma\mathbf{M}\times(\mathbf{H}%
_{\mathrm{app}}+\alpha_{\mathrm{ex}}\nabla^{2}\mathbf{M})+\frac{\alpha
_{\mathrm{G}}}{M_{s}}\mathbf{M}\times\frac{d\mathbf{M}}{dt},
\end{equation}
where $\alpha_{G}$ is the Gilbert damping constant and $-\gamma$ is the
electron gyromagnetic ratio. In the absence of external torques and damping
the magnetization carries a magnetization current density
\begin{equation}
\tilde{\mathbf{j}}_{\delta}=\alpha_{\mathrm{ex}}\mu_{0}\gamma\mathbf{M}%
\times\nabla_{\delta}\mathbf{M}, \label{spin_current_density}%
\end{equation}
which satisfies the continuity equation $d\mathbf{M}/dt+\nabla\cdot
\tilde{\mathbf{j}}=0$ \cite{Yaroslav_spin_current}. When considering the
excitation of magnetization, we include the microwave field $\mathbf{H}(t)$ in
the LLG equation.

The microwaves emitted by a long stripline on top of a thin magnetic film
launch a coherent magnon current normal to it. We consider a metallic wire of
rectangular cross section $0<x<d$ and $-w/2<y<w/2$ (Fig.~\ref{nonlinear_YIG})
with an AC current density $I$ of frequency $\omega_{s}$. The microwaves are
uniform over the film thickness when $s\ll d$. The Fourier component $k_{y}$
of the Oersted magnetic field in the thin film below the stripline
($x\rightarrow-s/2$) reads
\cite{Jackson,poineering_1,poineering_2,stripline_chirality,nano_optics,Toeno_NV}%
,
\begin{align}
H_{x}(k_{y},\omega_{s})  &  =(i/2)I(\omega_{s})\mathcal{F}%
(d,w)\mathrm{sgn}(k_{y})e^{-|k_{y}|(d+s)/2},\nonumber\\
H_{y}(k_{y},\omega_{s})  &  =-(1/2)I(\omega_{s})\mathcal{F}%
(d,w)e^{-|k_{y}|(d+s)/2}, \label{stripline_field}%
\end{align}
with $\mathcal{F}(d,w)=({2}/{k_{y}^{2}})\sin\left(  k_{y}{w}/{2}\right)
\left(  1-e^{-|k_{y}|d}\right)  $ determined by stripline dimensions. Here we
used $\left\vert k_{y}\right\vert \gg\omega_{s}/c$ because the velocity of
light $c$ is much larger than that of the magnons. The magnetic field
$H_{y}(k_{y},\omega_{s})=i\mathrm{sgn}(k_{y})H_{x}(k_{y},\omega_{s})$ is right
and left circularly polarized for positive and negative $k_{y}$, respectively,
so polarization and momentum are locked. In the linear regime, this field
coherently excites circularly-polarized magnons that propagate
unidirectionally and populate at one side of the stripline, i.e., a chiral
pumping effect. This picture will be thoroughly changed in the nonlinear
regime, however (see below).

Figure~\ref{fig:spin_current} illustrates the pumped magnon spin current
${\mathbf{J}}_{y}(y=0)=-1/(2\omega_{M}\gamma\alpha_{\mathrm{ex}})\int_{-s}%
^{0}dx\tilde{\mathbf{j}}_{y}(x,y=0)$ with $\omega_{M}\equiv\mu_{0}\gamma
M_{s}$ as a function of the applied electric current density $I$ with
frequency $\omega_{s}\approx\{5.8,11.3\}$~GHz across the stripline of width
$w=\{150,200\}$~nm and thickness $d=80$~nm \cite{7nm,Toeno_NV} from numerical
solutions of the LLG equation. Here the YIG film thickness $s=10$~nm, the
applied static magnetic field $\mu_{0}H_{\mathrm{app}}=10$~mT that drives out
domain walls \cite{7nm,one_wire}, $\mu_{0}M_{s}=0.18$~T, and $\alpha
_{G}=10^{-4}$. With the increase of the biased current in the stripline, the
spin current firstly linearly increase but become saturated or maximal at
a critical electric current $I_{c}$. This phenomenon is completely unexpected
for non-interacting magnons that should scale as $|\mathbf{J}_{y}|\propto I^{2}%
$, which highlights the importance of the interaction effects
discussed in the following. 

\begin{figure}[th]
{\includegraphics[width=5.6cm]{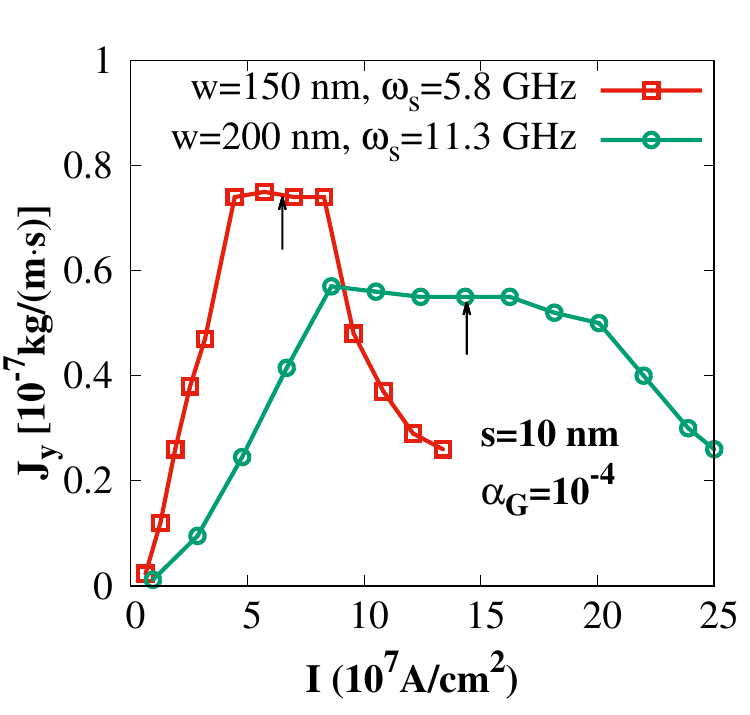}} \caption{(Color online)
Maximum spin current excited in a magnetic film with thicknss $s=10$ nm by an
AC charge current density $I$ in a proximity microwave stripline calculated by
numerically solving the LLG equation. The black arrows indicate the critical
current density $I_{c}$ for the indicated frequencies $\omega_s$ and stripline widths $w$.}%
\label{fig:spin_current}%
\end{figure}

\textit{Magnonic Doppler effect.}---The LLG phenomenology contains all of the
nonlinearities that can be captured by interacting magnons to some extent. The
Holstein-Primakoff transformation expresses the magnetization dynamics by
bosonic magnon operators $\hat{\Theta}(\mathbf{r})$ with $\hat{S}%
_{x}(\mathbf{r})+i\hat{S}_{y^{\prime}}(\mathbf{r})=\hat{\Theta}^{\dagger
}(\mathbf{r})\sqrt{2S-\hat{\Theta}^{\dagger}(\mathbf{r})\hat{\Theta
}(\mathbf{r})}$ and $\hat{S}_{z^{\prime}}(\mathbf{r})=-S+\hat{\Theta}%
^{\dagger}(\mathbf{r})\hat{\Theta}(\mathbf{r})$, where the spin operators
$\hat{\mathbf{S}}=-\mathbf{M/}\left(  \gamma\hbar\right)  $. The leading terms
in the expansion of the square roots leads to a complete set of harmonic
oscillators that we use to expand the full problem. The eigenmodes normal to
the film plane depend on the boundary conditions that become free for thin
films \cite{Wang2019}. The magnon operators in position space can then be
expanded in perpendicular standing spin waves (PSSWs) with index $l$
\cite{Bayer,chiral_excitation}
\begin{equation}
\hat{\Theta}(\mathbf{r})=\sum_{l\geq0}\sqrt{\frac{2}{1+\delta_{l0}}}\frac
{1}{\sqrt{s}}\cos\left(  \frac{l\pi}{s}x\right)  \hat{\Psi}_{l}%
({\boldsymbol{\rho}}),
\end{equation}
where ${\boldsymbol{\rho}}=y\hat{\mathbf{y}}+z\hat{\mathbf{z}}$. Substituting
these expressions into the Holstein-Primakoff expansion, the Hamiltonian can
be written as $\hat{H}=\hat{H}_{\mathrm{L}}+\hat{H}_{\mathrm{NL}}+\cdots$,
where $\hat{H}_{\mathrm{L}}$ describes the non-interacting magnon gas and
$\hat{H}_{\mathrm{NL}}$ is the leading nonlinear term that introduces
interactions between the magnons. At sufficiently low magnon densities
\begin{equation}
\hat{H}\rightarrow\hat{H}_{\mathrm{L}}=\sum_{l}(E_{l}+\hbar\omega_{M}%
\alpha_{\mathrm{ex}}k^{2})\int\hat{\Psi}_{l}^{\dagger}({\boldsymbol{\rho}%
})\hat{\Psi}_{l}({\boldsymbol{\rho}})d{\boldsymbol{\rho}},
\end{equation}
where $E_{l}=\mu_{0}\gamma\hbar H_{\mathrm{app}}+\hbar\omega_{M}%
\alpha_{\mathrm{ex}}(l\pi/s)^{2}$ is the edge of the $l$-th band. The
nonlinear Hamiltonian
\begin{align}
&  \hat{H}_{\mathrm{NL}}=\sum_{l_{i}}\mathcal{U}_{l_{1}l_{2}l_{3}l_{4}}%
\int\hat{\Psi}_{l_{1}}^{\dagger}({\boldsymbol{\rho}})\hat{\Psi}_{l_{2}%
}^{\dagger}({\boldsymbol{\rho}})\hat{\Psi}_{l_{3}}({\boldsymbol{\rho}}%
)\hat{\Psi}_{l_{4}}({\boldsymbol{\rho}})d{\boldsymbol{\rho}}\nonumber\\
&  +\sum_{l_{i}}\mathcal{V}_{l_{1}l_{2}l_{3}l_{4}}\int\hat{\Psi}_{l_{1}%
}^{\dagger}({\boldsymbol{\rho}})\hat{\Psi}_{l_{2}}^{\dagger}({\boldsymbol{\rho
}})\nabla_{\boldsymbol{\rho}}\hat{\Psi}_{l_{3}}\cdot\nabla_{\boldsymbol{\rho}%
}\hat{\Psi}_{l_{4}}d{\boldsymbol{\rho}}+\mathrm{H.c.}\nonumber
\end{align}
contains two types of magnon-number conserving interactions derived in the
Supplemental Material \cite{supplement}. The potentials 
\begin{align}
\mathcal{U}_{l_{1}l_{2}l_{3}l_{4}}  &  =\frac{\mu_{0}\gamma^{2}\hbar^{2}%
\alpha_{\mathrm{ex}}l_{3}l_{4}\pi^{2}\mathcal{A}_{l_{1}l_{2}l_{3}l_{4}}}%
{s^{3}\sqrt{(1+\delta_{l_{1}0})(1+\delta_{l_{2}0})(1+\delta_{l_{3}0}%
)(1+\delta_{l_{4}0})}},\nonumber\\
\mathcal{V}_{l_{1}l_{2}l_{3}l_{4}}  &  =\frac{\mu_{0}\gamma^{2}\hbar^{2}%
\alpha_{\mathrm{ex}}\mathcal{B}_{l_{1}l_{2}l_{3}l_{4}}}{s\sqrt{(1+\delta
_{l_{1}0})(1+\delta_{l_{2}0})(1+\delta_{l_{3}0})(1+\delta_{l_{4}0})}%
},\nonumber
\end{align}
are governed by
magnon-mode overlap integrals
\begin{align}
\mathcal{A}_{l_{1}l_{2}l_{3}l_{4}}  &  =\frac{1}{s}\int_{-s}^{0}dx\Pi
_{i=1,2}\cos\left(  \frac{l_{i}\pi}{s}x\right)  \Pi_{j=3,4}\sin\left(
\frac{l_{j}\pi}{s}x\right)  ,\nonumber\\
\mathcal{B}_{l_{1}l_{2}l_{3}l_{4}}  &  =\frac{1}{s}\int_{-s}^{0}%
dx\Pi_{i=1,2,3,4}\cos\left(  \frac{l_{i}\pi}{s}x\right)  .\nonumber
\end{align}
When $l_{1}=0$, the scattering potentials obey selection rules $\mathcal{U}%
_{0l_{2}l_{3}l_{4}}\propto l_{3}l_{4}\left(  \delta_{l_{2}+l_{3},l_{4}}%
+\delta_{l_{2}+l_{4},l_{3}}-\delta_{l_{3}+l_{4},l_{2}}\right)  $ and
$\mathcal{V}_{0l_{2}l_{3}l_{4}}\propto\left(  \delta_{l_{2}+l_{3},l_{4}%
}+\delta_{l_{2}+l_{4},l_{3}}+\delta_{l_{2}+l_{3}+l_{4},0}+\delta_{l_{3}%
+l_{4},l_{2}}\right)  $. In the two-dimensional limit, $\mathcal{U}%
_{0000}=0$ vanishes, but $\mathcal{V}_{0000}=\mathcal{V}_{00ll}=\mathcal{V}%
_{0}=\mu_{0}\gamma^{2}\hbar^{2}\alpha_{\mathrm{ex}}/\left(  4s\right)  $ is
large. The divergence for vanishing film thickness is an artifact of the
continuum approximation that breaks down when $s$ approaches unit cell dimensions.

We are interested in the effect of a magnon current on a low-frequency
coherent excitation, i.e., at excitation frequency $\omega/(2\pi)\lesssim
1$~GHz, which allows us to set $l_{1}=0$. Using the above selection rules of
the scattering potentials and energy conservation, we prove in the
Supplemental Material \cite{supplement} that the incoherent scattering of
these \textit{low-energy} magnons by those in all other bands is marginally
small. The leading nonlinearities in the coherent magnon states thus reduce to
a self-consistent mean-field problem \cite{Abrikosov,ZNG}, in which the
interaction renormalizes the energy dispersion but does not affect magnon
dephasing and lifetime. The coherent magnon amplitude in the lowest band obeys
a Heisenberg equation of motion that is augmented by the Gilbert damping
\cite{supplement},
\begin{align}
&  i\hbar(1-i\alpha_{G})\frac{\partial\langle\hat{\Psi}_{0}(\boldsymbol{\rho
})\rangle}{\partial t}=E_{0}\langle\hat{\Psi}_{0}(\boldsymbol{\rho}%
)\rangle-\hbar\omega_{M}\alpha_{\mathrm{ex}}\nabla^{2}\langle\hat{\Psi}%
_{0}(\boldsymbol{\rho})\rangle\nonumber\\
&  +\frac{8i}{\hbar}\sum_{l^{\prime}\geq0}{\mathcal{V}}_{00l^{\prime}%
l^{\prime}}\mathbf{J}_{l^{\prime}}({\boldsymbol{\rho}})\cdot\nabla
_{\boldsymbol{\rho}}\langle\hat{\Psi}_{0}({\boldsymbol{\rho}})\rangle
+P_{\mathrm{ex}}, \label{EOM_all}%
\end{align}
where $\langle\cdots\rangle$ represents an ensemble average,
\begin{equation}
\mathbf{J}_{l}({\boldsymbol{\rho}})=\frac{\hbar}{2i}\left(  \langle\hat{\Psi
}_{l}^{\dagger}({\boldsymbol{\rho}})\nabla_{\boldsymbol{\rho}}\hat{\Psi}%
_{l}({\boldsymbol{\rho}})\rangle-\langle\hat{\Psi}_{l}({\boldsymbol{\rho}%
})\nabla_{\boldsymbol{\rho}}\hat{\Psi}_{l}^{\dagger}({\boldsymbol{\rho}%
})\rangle\right)  \label{magnon_current}%
\end{equation}
is the magnon linear-momentum current density in subband $l$ with
contributions from both coherent and incoherent magnons, and $P_{\mathrm{ex}}$
is a microwaves excitation source that will be specified below. The (locally)
uniform magnon current hence engages the gradient (or momentum) of the magnon
amplitude $\nabla_{\boldsymbol{\rho}}\langle\hat{\Psi}_{0}\rangle$ and tilts
the magnon dispersion, which is an interaction-induced drag effect
\cite{phonon,electron}.

The magnon momentum current density [Eq.~(\ref{magnon_current})] is
proportional to the magnon-number current density $\mathbf{\tilde{J}}_{l}$
defined by the continuity and Heisenberg equations for
the non-interacting magnon Hamiltonian, since the exchange magnons have a
constant mass $\hbar/(2\omega_{M}\alpha_{\mathrm{ex}})$. The former is also a
spin current since in the absence of anisotropies the magnons carry angular
momentum $\hbar$. With magnon density operator $\hat{\rho}_{m}^{l}%
({\boldsymbol{\rho}})=\langle\hat{\Psi}_{l}^{\dagger}({\boldsymbol{\rho}}%
)\hat{\Psi}_{l}({\boldsymbol{\rho}})\rangle$
\begin{equation}
\frac{\partial\hat{\rho}_{m}^{l}({\boldsymbol{\rho}})}{\partial t}=\frac
{1}{i\hbar}[\hat{\rho}_{m}^{l}({\boldsymbol{\rho}}),\hat{H}_{\mathrm{L}%
}]=-\nabla\cdot\mathbf{\tilde{J}}_{l}({\boldsymbol{\rho}}),
\label{LinearHeisen}%
\end{equation}
leading to $\left\langle \mathbf{\tilde{J}}_{l}({\boldsymbol{\rho}%
})\right\rangle =(2{\omega_{M}\alpha_{\mathrm{ex}}}/{\hbar})\mathbf{J}%
_{l}({\boldsymbol{\rho}}),$ which is consistent with
Eq.~(\ref{spin_current_density}) since $-1/(\gamma\hbar)\int dx\tilde
{\mathbf{j}}(x,{\boldsymbol{\rho}})\rightarrow\tilde{\mathbf{J}}%
_{l}({\boldsymbol{\rho}})$ when $l=0$ to linear order in the magnon operator.

This stripline microwave field [Eq.~(\ref{stripline_field})] couples to the
magnons of the lowest PSSW band up to wave numbers $k_{y}\sim\pi/w$ by the
Zeeman interaction
\[
\hat{H}_{\mathrm{Z}}=g\sum_{k_{y}}\left(  H_{x}(k_{y},t)-i\cos\varphi
H_{y}(k_{y},t)\right)  \hat{\Psi}_{0}^{\dagger}(k_{y})+\mathrm{H.c.},
\]
with coupling constant $g=\mu_{0}\sqrt{\gamma\hbar M_{s}s/2}$, so the
excitation source $P_{\mathrm{ex}}=g(H_{x}(k_{y},t)-i\cos\varphi H_{y}%
(k_{y},t))$ in Eq.~(\ref{EOM_all}). The in-plane magnetization angle $\varphi$
can be rotated by an applied DC magnetic field to tune the magnitude and
direction of the pumped magnon current. When $\varphi=0$, the stripline
magnetic field launches a magnon current with $k_{y}>0$ into half space (see
below). Thereby the excited magnon current $\mathbf{J}_{y}(y>0)=\overline
{\mathbf{J}}_{y}\exp({-y/\delta})$ decays exponentially with distance from the
source on the scale of the decay length $\delta\left(  \omega_{s}\right)
\sim2/\operatorname{Im}\kappa_{y}\sim\sqrt{(\alpha_{\mathrm{ex}}\omega
_{M})(\omega_{s}-\mu_{0}\gamma H_{\mathrm{app}})}/(\alpha_{G}\omega_{s})$,
i.e. the root of $(\omega_{s}-\mu_{0}\gamma H_{\mathrm{app}}-\omega_{M}%
\alpha_{\mathrm{ex}}\kappa_{y}^{2})^{2}+(\alpha_{G}\omega_{s})^{2}=0$. On the
other hand, the amplitude $\langle\hat{\Psi}_{0}({\boldsymbol{\rho}})\rangle$
oscillates rapidly with wavelength ($1/|\kappa_{y}|\ll\delta$). Near the
stripline, the magnon current in the lowest band obeys the integral equation,
obtained from Eq.~(\ref{EOM_all}),
\begin{equation}
\overline{\mathbf{J}}_{y}=\frac{1}{\delta}\left(  \frac{g}{\hbar}\right)
^{2}\int\frac{dk_{y}}{2\pi}k_{y}\frac{\left\vert H_{x}(k_{y})-i H_{y}%
(k_{y})\right\vert ^{2}}{(\omega_{s}-\tilde{\omega}_{k_{y}})^{2}+\alpha
_{G}^{2}\omega_{s}^{2}}, \label{stripline_spin_current}%
\end{equation}
with Doppler-shifted magnon frequency
\begin{equation}
\tilde{\omega}_{\mathbf{k}}=\mu_{0}\gamma H_{\mathrm{app}}+\omega_{M}%
\alpha_{\mathrm{ex}}k^{2}-({8}/{\hbar^{2}}){\mathcal{V}}_{0}k_{y}%
\overline{\mathbf{J}}_{y}, \label{excitations}%
\end{equation}
which can be solved iteratively or graphically.

\begin{figure}[th]
\hspace{-0.19cm}{\includegraphics[width=4.43cm]{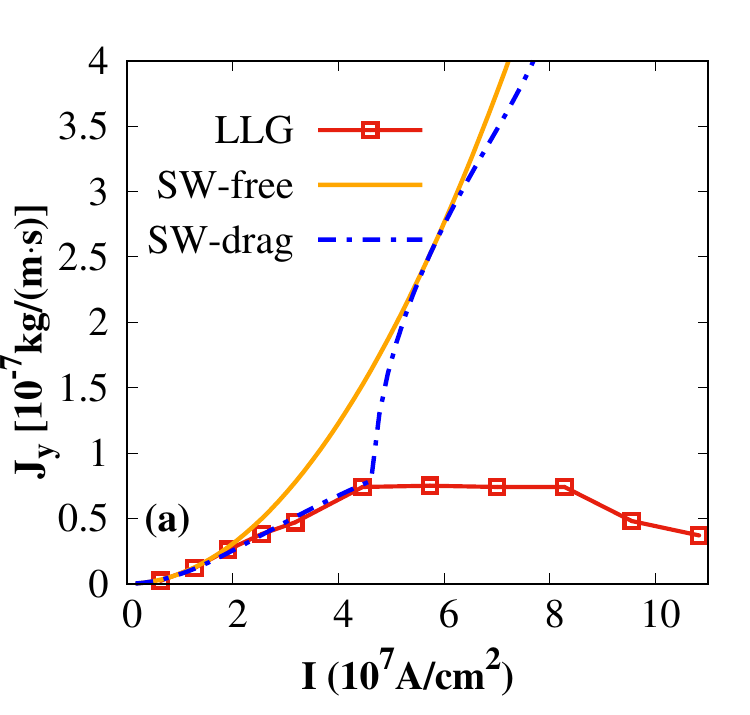}}
\hspace{-0.35cm}{\includegraphics[width=4.43cm]{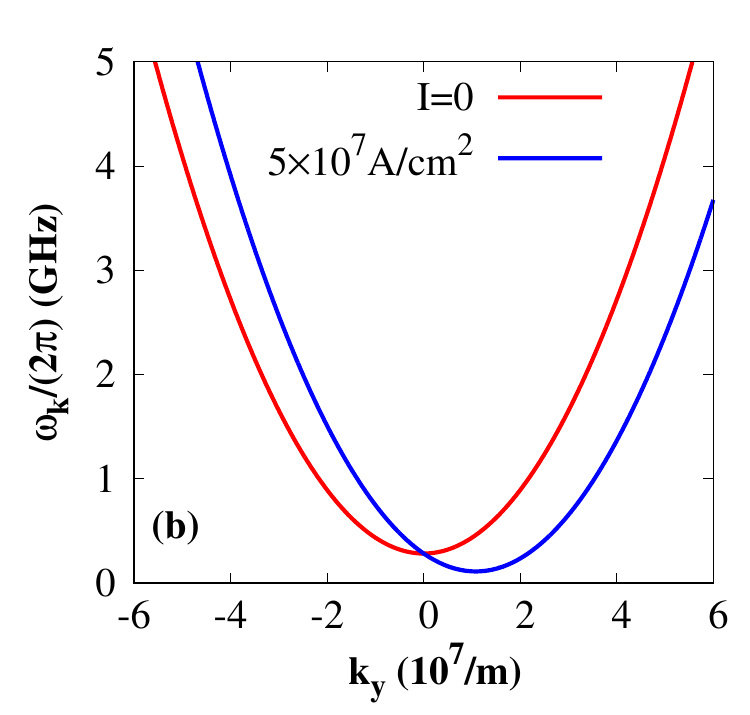}}
\newline\hspace{-0.37cm}{\includegraphics[width=4.44cm]{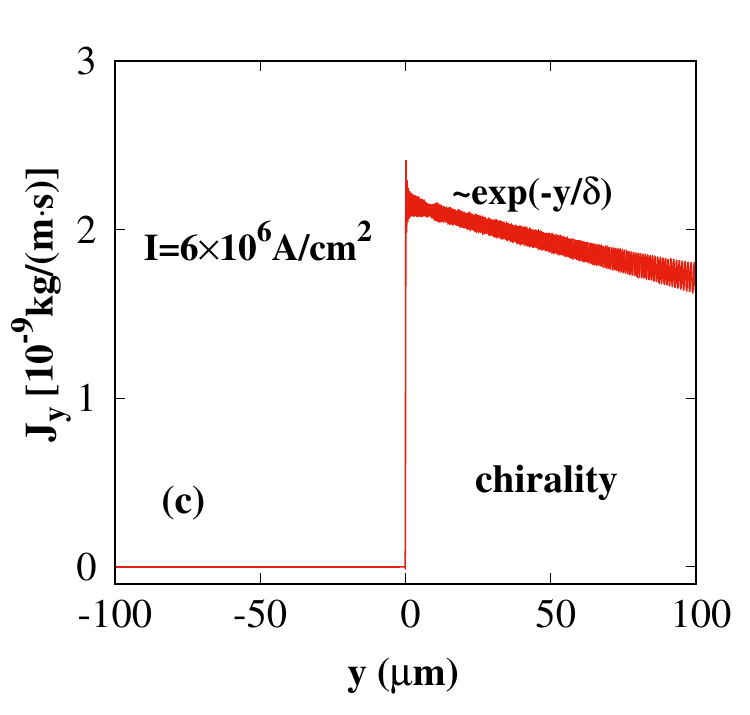}}
\hspace{-0.33cm}{\includegraphics[width=4.44cm]{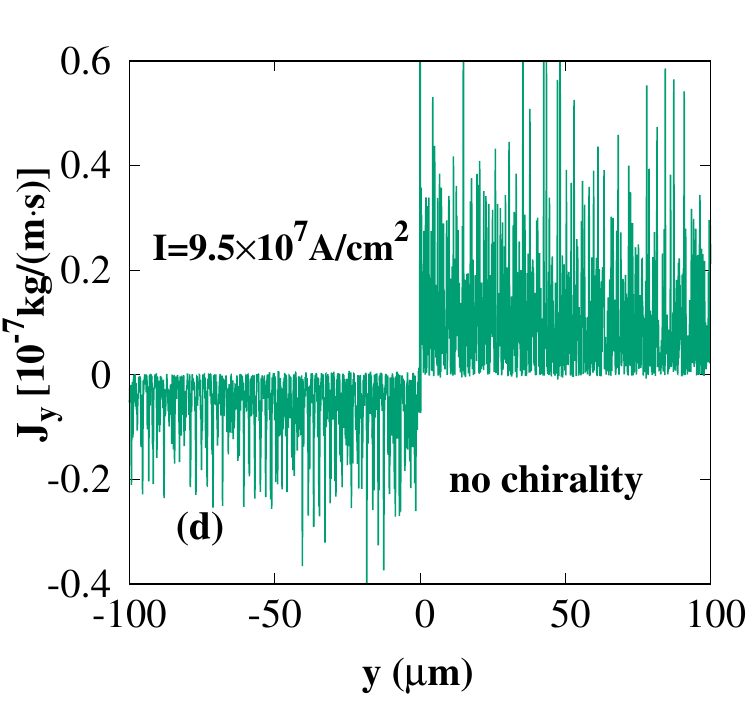}}
\caption{(Color online) Magnon currents and Doppler shift of the magnon
dispersion under stripline microwave excitation. (a) shows the coherently
pumped magnon current $\overline{\mathbf{J}}_{y}$ as a function of the applied
electric current density $I$ in the stripline from numerical LLG calculations
(\textquotedblleft LLG"), non-interacting spin-wave theory (\textquotedblleft
SW-free"), and spin-wave theory including the drag effect (\textquotedblleft
SW-drag"). The tilt of the magnon dispersion at high excitation is illustrated
in (b). We illustrate the chirality of the spin-current excitation for
$I<I_{c}$ [(c)] and $I>I_{c}$ [(d)], respectively.}%
\label{fig:shift}%
\end{figure}

\begin{figure}[th]
\hspace{-0.15cm}{\includegraphics[width=4.44cm]{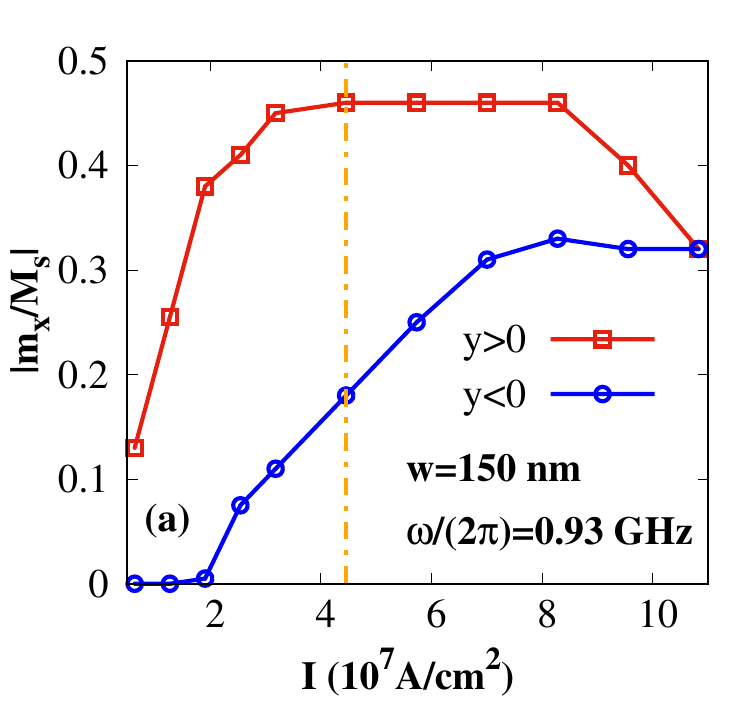}}
\hspace{-0.18cm}{\includegraphics[width=4.44cm]{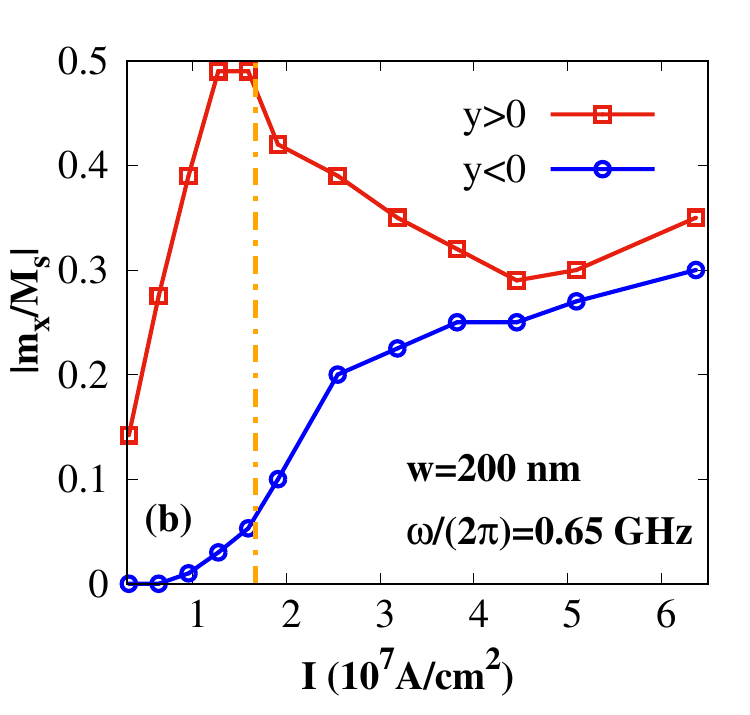}}
\caption{(Color online) Magnon densities (reduced magnetization) at the right
and left side of a stripline as a function of current density $I$ with
excitation frequencies $\omega_{s}=2\pi\times0.93$~GHz, width $w=150$ nm [(a)]
and $2\pi\times0.65$~GHz width $w=200$ nm [(b)]. The vertical orange line
indicates the critical $I_{c}$ that maximizes the spin current.}%
\label{fig:amplitude}%
\end{figure}

Figure~\ref{fig:shift}(a) illustrates the pumped magnon current $\overline
{\mathbf{J}}_{y}$ as a function of the applied electric current density $I$
with frequency $\omega_{s}/(2\pi)\approx0.93$~GHz across the stripline of
width $w=150$~nm and thickness $d=80$~nm \cite{7nm,Toeno_NV} from
Eq.~(\ref{stripline_spin_current}) in comparison with numerical solutions of
the LLG equation [Eq.~(\ref{spin_current_density})]. 
Magnons of wavelength $2w$ are resonantly excited and carry a
current with decay length $\delta\approx333~\mathrm{\mu}$m. Here we compare
the analytical solutions with the numerically exact solution of the LLG
equation, which predicts a maximum spin-wave current for a stripline current
$I_{c}\approx5\times10^{7}~\mathrm{A/cm^{2}}$. The non-interacting spin-wave
theory (SW-free) fails already for small $I$, which emphasizes the importance
of nonlinearities. When including the drag effect, the spin-wave theory
Eq.~(\ref{stripline_spin_current}) $\overline{\mathbf{J}}_{y}$ saturates at a
current $I\sim I_{c}$, but returns to the non-interacting values at larger
currents. When $I>I_{c}$, the lowest-order nonlinearity of the
Holstein-Primakoff expansion and thereby the mean-field theory may break down.
The Doppler shift of the spin-wave dispersion illustrated in
Fig.~\ref{fig:shift}(b) holds only for $I<I_{c}$. More detailed comparison
with different parameters confirms these features \cite{supplement}. When
$I\gtrsim I_{c}$, we observe that the chirality of the magnon excitation is
strongly reduced, indicating that the backscattering of magnons becomes
strong, as illustrated by Figs.~\ref{fig:amplitude}(c) and (d), which is
partly responsible for the suppression of spin current.

$I_{c}$ can be estimated by the onset of a spin-wave instability that is
characterized by negative magnon excitation energy
\cite{instability_Doppler1,instability_Doppler2,Tatara}, which causes the
discontinuous change of the spin current calculated by the
mean-field theory. According to Eq.~(\ref{excitations}) a critical magnon
current
\begin{equation}
\mathbf{J}_{y}^{\left(  c\right)  }=\hbar/(4\mathcal{V}_{0})\sqrt{\hbar
\omega_{M}\alpha_{\mathrm{ex}}E_{0}}%
\end{equation}
can cause negative magnon excitation energies $\tilde{E}_{0}(\mathbf{k})<0$ at
the momentum $k_{y}^{(c)}=4\mathcal{V}_{0}\overline{\mathbf{J}}_{y}/(\hbar
^{2}\omega_{M}\alpha_{\mathrm{ex}})$. With the above YIG parameters, the
critical magnon current $\mathbf{J}_{y}^{(c)}\approx10^{-7}~\mathrm{kg/(m\cdot
s)}$. This value can be reached by incoherent spin injection with a critical
temperature gradient $4\,\mathrm{K/\mu m}$ when $T=300$~K \cite{supplement}.
However, according to the LLG calculations in Fig.~\ref{fig:spin_current} with
different material parameters nonlinearities might prohibit reaching this
critical value, which thus provides a upper limit in the estimation of maximal
spin currents (more detailed comparison refers to the Supplemental Material
\cite{supplement}).

The tilt of dispersion causes chiral velocities of spin waves of the same
energy that should be observable by changes in the microwave transmission
\cite{7nm,one_wire}, nitrogen-vacancy center magnetometry
\cite{Toeno_review,Toeno_NV}, and Brillouin light scattering \cite{BLS_review}%
. The dispersion tilts into the opposite direction when the magnetization
direction is reversed ($\varphi=\pi$) and vanishes when perpendicular to the
stripline ($\varphi=\pi/2$), i.e., it follows the current direction governed
by the chirality of the stripline magnetic field. The basic features agree
with recently reported experiments in YIG thin films of thickness $s=7$~nm
\cite{7nm} that were interpreted in terms of the DMI although spin-orbit
interaction is small for closed-shell magnetic moments \cite{Caretta}. The
Doppler effect, on the other hand, is tunable by the magnitude and direction
of the excited magnonic spin current and does not require special interfaces.
We note that an interficial DMI causes additional shift of the magnon
dispersion that favors the realization of spin-wave instability as calculated
in the Supplemental Material \cite{supplement}.

\textit{Breaking of chiral pumping.}---Finally, mean-field theory reveals a
connection between the breakdown of the chiral pumping and the spin-wave
instability. Around the critical driving strength $I_{c}$ the magnon density
on one side of the stripline reaches its maximum with a rapid increase of the
magnon density on the other side. Figure~\ref{fig:amplitude} shows the
suppression of chirality under strong excitation. The nonequilibrium
magnetization for $y>0$ is largest around $I_{c}$, at which magnons accumulate
also at $y<0$. The chirality is strongly broken for larger drives, with nearly
equal excited magnon densities on both sides of the stripline such that the
injected power propagates into both directions, similar to the electric or
thermal injection of an incoherent magnon accumulation.

\textit{Discussion.}---In conclusion, we formulated the dynamics of a strongly
driven ultrathin film of magnetic insulator such as YIG. We predict a
Doppler shift of the magnon dispersion and a maximum spin current that a given
sample can sustain. In our example, the effects should occur at stripline
current densities $\sim2\times10^{7}~\mathrm{A/cm^{2}}$ in one or
$\sim(2/\mathcal{N})\times10^{7}~\mathrm{A/cm^{2}}$ in $\mathcal{N}$
striplines (distributed over a total width that should be small compared to
the magnon propagation length, i.e., many micrometers). The nonmonotonic
dependence of the spin current excited by microwaves power may be related to
the observed non-monotonicity of spin transport in magnon transistors as a
function of gate-injected magnon densities
\cite{nonlinear_electric1,nonlinear_electric2}. Our theory should help
understanding the effects of large magnon spin currents on the magnetic order
of insulators and provides a different scenario for the nonlinearities induced
by the magnon chemical potential
\cite{Suhl,turbulence,magnon_BEC,nonlinear_microwave,SHE_BEC1,BEC_Flebus,chemical_NV,Navier_Stokes}%
.

\vskip0.25cm \begin{acknowledgments}
This work is financially supported by DFG Emmy Noether program (SE 2558/2-1) as well as JSPS KAKENHI Grant No. 19H006450.  C.W. is supported by the National Natural Science Foundation of China under Grant No. 11704061.
We thank  Xiang-Yang Wei, Hanchen Wang, Haiming Yu, and Mehrdad Elyasi for valuable discussions.
\end{acknowledgments}

\end{document}